\documentclass[preprint,superscriptaddress,floatfix,preprintnumbers,nofootinbib,altaffilletter]{revtex4-1}
\usepackage{graphicx}
\usepackage{subfigure}
\usepackage{dcolumn}
\usepackage{bm}
\usepackage{amssymb}
\usepackage{xcolor}
\usepackage[utf8]{inputenc}
\usepackage[OT1]{fontenc}
\usepackage{yfonts,amsmath,amsthm,amsfonts,amssymb,amscd, ulem}
\usepackage{enumerate}
\usepackage{fancyhdr}
\usepackage{mathtools}
\usepackage{mathrsfs}
\usepackage{cancel}
\usepackage{slashed}
\usepackage{bigints}
\usepackage[flushleft]{threeparttable}
\usepackage[colorlinks=true, citecolor=purple, linkcolor=blue]{hyperref}
\usepackage{url}
\usepackage{makecell,booktabs}
\usepackage{braket}
\usepackage{relsize}
\usepackage{multirow}
\usepackage{verbatim}
\usepackage{txfonts}
\usepackage{upgreek}
\usepackage{extarrows}
\usepackage{array}
\usepackage{appendix}
\usepackage[T1]{fontenc}
\usepackage{setspace}

\renewcommand{\arraystretch}{0.6}

\begin{document}
\title{Long-lived Sterile Neutrino Searches at Future Muon Colliders}

\author{Qi Bi}
\email{biqii@buaa.edu.cn}
\affiliation{School of Physics, Beihang University, Beijing 100083, China}

\author{Jinhui Guo}
\email{guojh23@buaa.edu.cn}
\affiliation{School of Physics, Beihang University, Beijing 100083, China}

\author{Jia Liu}
\email{jialiu@pku.edu.cn}
\affiliation{School of Physics and State Key Laboratory of Nuclear Physics and Technology, Peking University, Beijing 100871, China}
\affiliation{Center for High Energy Physics, Peking University, Beijing 100871, China}

\author{Yan Luo}
\email{ly23@stu.pku.edu.cn}
\affiliation{School of Physics and State Key Laboratory of Nuclear Physics and Technology, Peking University, Beijing 100871, China}

\author{Xiao-Ping Wang}
\email{hcwangxiaoping@buaa.edu.cn}
\affiliation{School of Physics, Beihang University, Beijing 100083, China}
\affiliation{Beijing Key Laboratory of Advanced Nuclear Materials and Physics, Beihang University, Beijing 100191, China}

\begin{abstract}

We explore the potential of studying sterile neutrinos at a future high-energy muon collider, where these particles can generate small active neutrino masses via the seesaw mechanism and exhibit long-lived particle signatures. A Dirac sterile neutrino model with ${\rm U(1)}_{L_\mu-L_\tau}$ symmetry is introduced, where the heavy right-handed neutrino ($N_R$) produces tiny active neutrino masses, and the light left-handed neutrino ($N_L$) naturally behaves as a long-lived particle. The ${\rm U(1)}_{L_\mu-L_\tau}$ gauge symmetry also enhances sterile neutrino pair production at a future high-energy muon collider. Using the displaced vertex method, the muon collider can search for heavy sterile neutrino, especially for $m_L> m_W$. We find that a muon collider with $\sqrt{s} = 3~ (10)$ TeV and luminosity $\mathcal{L}=1~(10)$ ab$^{-1}$ can probe $N_L$ masses of $m_L \in [100,~1500~(5000)]$ GeV and mixing angles $\theta_{\nu L} \in [10^{-13},~10^{-6}]$.

\end{abstract}

\maketitle
\tableofcontents

\newpage
\section{Introduction}
\label{sec:int}

Neutrino oscillation experiments have shown that at least two generations of neutrinos have small but nonzero masses \cite{Davis:1968cp,KamLAND:2002uet,DayaBay:2012fng,DoubleChooz:2014kuw}, contradicting the Standard Model (SM), which assumes neutrinos are massless. This discrepancy, known as the neutrino mass puzzle, provides a key opportunity to investigate physics beyond the Standard Model (BSM). A natural solution to this puzzle is offered by the seesaw mechanism, which predicts the existence of a heavy right-handed neutrino that mixes with Standard Model neutrinos, giving rise to their tiny masses \cite{Minkowski:1977sc,Yanagida:1979as,Mohapatra:1979ia,Gell-Mann:1979vob}. Numerous experiments have diligently searched for sterile neutrinos, spanning both oscillation experiments and collider studies \cite{Kopp:2013vaa,Giunti:2015wnd,Keung:1983uu,Han:2006ip,delAguila:2007qnc,Cai:2017mow,Accomando:2017qcs,Cvetic:2018elt,Das_2019}. 

There are plenty of seesaw models \cite{Schechter:1980gr,Schechter:1981cv,Akhmedov:1995ip,Akhmedov:1995vm,Barr:2003nn,Wyler:1982dd,Mohapatra:1986bd,Mohapatra:1986aw,Das:2012ze}, which can generate the tiny neutrino mass via a heavy right-handed neutrino. In the type-I seesaw mechanism, if the Dirac mass from the Higgs mechanism is at the MeV scale, the typical sterile neutrino mass is around $m_{\nu_s} \sim 100$ TeV. The corresponding mixing angle between active and sterile neutrinos is constrained by the upper limits on the neutrino mass to approximately $\theta \sim \sqrt{m_\nu/m_{\nu_s}} \sim 10^{-8}$ \cite{Planck:2018vyg,DiValentino:2021hoh}. However, the heavy mass and extremely small couplings in the canonical type-I seesaw framework make detecting sterile neutrinos challenging. Besides, in other mechanisms, such as linear and inverse seesaw mechanisms \cite{Akhmedov:1995ip,Akhmedov:1995vm,Mohapatra:1986aw}, the sterile neutrinos can be light with mixing parameters larger than the values required by the type-I seesaw mechanism, while still explaining the tiny active-neutrino masses. As a result, researchers often treat sterile neutrino properties as free parameters when exploring collider phenomenology. 

Due to the large QCD backgrounds and the small mixing, searching for sterile neutrinos at LHC and HL-LHC with long-lived signatures helps to increase the sensitivity \cite{Drewes:2019vjy,Dib:2019ztn,Drewes:2019fou,Bondarenko:2019tss,Drewes:2019vjy,Beltran:2021hpq,Cottin:2021lzz,CMS:2024hik,CMS:2024ake}. There have been several studies on searching for long-lived sterile neutrinos that are lighter than gauge bosons at LHC \cite{Helo:2013esa,Antusch:2017hhu,Helo:2018qej,Bondarenko:2019tss,Liu:2019ayx,CMS:2022fut,CMS:2024hik}, where their decays are mainly to three-body final states at this mass range. Moreover, the future high-energy muon collider presents a significant opportunity to explore high-energy scale physics at the TeV level due to its high energy, high luminosity, and reduced background \cite{Skrinsky:1981ht,Neuffer:1983xya,Neuffer:1986dg,Barger:1995hr,Barger:1996jm,Ankenbrandt:1999cta,Boscolo:2018ytm,Delahaye:2019omf,Long:2020wfp,AlAli:2021let,Aime:2022flm,Black:2022cth,Bose:2022obr,Narain:2022qud,Accettura:2023ked}. Many studies have investigated the muon collider potential to probe various aspects of both the precision measurements of the SM and BSM \cite{Han:2020pif,Yin:2020afe,Liu:2021jyc,Co:2022bqq,Ruhdorfer:2023uea,Liu:2023yrb,Li:2024joa,Cassidy:2023lwd}. The muon collider also opens new avenues for muon-specific research with this novel facility. There have been works searching for promptly decaying sterile neutrinos at the muon colliders in recent years \cite{Liu:2021akf,Li:2022kkc,Chakraborty:2022pcc,Mikulenko:2023ezx,Wang:2023zhh,Mekala:2023diu,Kwok:2023dck,Li:2023tbx,Li:2023lkl,Barducci:2024kig,He:2024dwh}, while their long-lived signatures are still absent. 

Addressing both the neutrino mass and long-lived sterile neutrino signatures at a muon collider is challenging. A promising solution involves introducing sterile neutrinos charged under the ${\rm U(1)}_{L_\mu-L_\tau}$ model \cite{He:1990pn,Cox:2017eme,Asai:2018ocx,Borah:2021mri,He:2024dwh}, which also helps resolve the muon $g-2$ anomaly. Recent studies~\cite{He:2024dwh} have examined the behavior of heavy neutral leptons (HNLs) in this framework, but focusing on the prompt decays of right-handed HNLs. Additionally, research on the ${\rm U(1)}_{L_\mu-L_\tau}$ gauge boson $Z'$ has primarily investigated processes like $\mu^- \mu^+ \to \gamma Z'$ with $Z' \to \mu^- \mu^+$ (or $\tau^- \tau^+)$, and $\mu^- \mu^+ \to \mu^+ \mu^-$ (or $\tau^+ \tau^-)$ mediated by $Z'$ \cite{Huang:2021nkl,Das:2022mmh}, focusing on invariant mass reconstruction or modifications to the SM processes to constrain the ${\rm U(1)}_{L_\mu-L_\tau}$ gauge coupling $g_{Z'}$.

In contrast to previous studies, we investigate a UV-complete model featuring a Dirac sterile neutrino charged under ${\rm U(1)}_{L_\mu-L_\tau}$, which subsequently splits into two HNLs: a heavy $N_R$ with a mass in the hundreds of TeV range, and a lighter $N_L$ with a mass greater than 100 GeV. The right-handed $N_R$ generates the small mass of SM neutrinos, similar to the type-I seesaw mechanism, while the left-handed $N_L$ is naturally long-lived due to the double suppression from the small neutrino mass and the small Dirac mass of sterile neutrinos. We explore the potential of detecting long-lived $N_L$ at a future high-energy muon collider with $\sqrt{s} = 3$ TeV and 10 TeV, focusing on its pair production through Drell-Yan processes mediated by the new $Z'$ gauge boson.  We then explore its subsequent long-lived decays into $Z \nu$, $W^\pm \ell^\mp$, or $h \nu$, and employ an inclusive search for displaced vertex signatures at the muon collider under specific benchmark parameter settings. Our results indicate strong sensitivity to the sterile neutrino parameter space, complementing other ongoing searches.

We organize the paper as follows. In section~\ref{sec:model}, we describe the model with two Majorana sterile neutrinos in gauged ${\rm U(1)}_{L_\mu-L_\tau}$ and their possible decay channels. In section~\ref{sec:constr}, we discuss the existing constraints from collider searches, neutrino trident process, and muon $g-2$ measurements. In section \ref{sec:prod}, we consider the possible production channels at the muon collider. In section~\ref{sec:pheno}, we discuss the long-lived particle (LLP) signatures and their detection at the muon collider. In section~\ref{sec:concl}, we conclude.

\section{The Model}
\label{sec:model}
To construct a UV-complete model, we extend the SM gauge group with an additional ${\rm U(1)}_{L_\mu-L_\tau}$ gauge group. Additionally, we introduce a new fermion $N$ and a complex scalar $\phi$, both of which are singlets under the SM group but charged under the ${\rm U(1)}_{L_\mu-L_\tau}$ gauge group. The gauge charges of these particles are summarized in Tab. \ref{tab:1}.

\begin{table}[ht]
\centering
\renewcommand{\arraystretch}{1.3}
\begin{tabular}{>{\centering}m{0.15\textwidth}|>{\centering}m{0.15\textwidth}|>{\centering}m{0.1\textwidth}|>{\centering}m{0.15\textwidth} |>{\centering}m{0.1\textwidth}|>{\centering}m{0.1\textwidth} | >{\centering\arraybackslash}m{0.1\textwidth} }
\hline
Gauge  Group & $L_\mu=\left (\begin{array}{c} \nu_{\mu,L} \\ \mu_L \end{array}\right)$ &  $\mu_R$ & $L_\tau=\left (\begin{array}{c} \nu_{\tau,L} \\ \tau_L \end{array}\right)$ &  $\tau_R$  &  $N$ &  $\phi$\\ \hline
 ${\rm SU(2)}_L$   & 2    & 1   & 2 & 1 & 1  & 1  \\ \hline
${\rm U(1)}_Y$ & -1 &-2 & -1 & -2 &0 & 0  \\ \hline
${\rm U(1)}_{L_\mu-L_\tau}$ & 1 & 1  & -1  & -1 &1 & -2 \\ \hline
\end{tabular}
\caption{Gauge charges of new particles and relevant SM particles in gauge groups.}
\label{tab:1}
\end{table}

Based on the charges of the particles under the gauge groups, the effective Lagrangian can be expressed as
\begin{equation}\label{eq:1}
\begin{aligned}
    \mathcal{L} \supset& -\frac{1}{4}Z'_{\mu\nu}Z'^{\mu\nu} + \sum_{\alpha=\mu, \tau} \left( i\bar{L}^0_\alpha \slashed{D}L_\alpha^0 + i\bar{\ell}_{\alpha,R}^0 \slashed{D}\ell_{\alpha,R}^0 \right) \\
    & + \bar{N}^0i\slashed{D}N^0 - m_N \bar{N}^0 N^0
    -y'\bar{L}^0_{\mu}\Tilde{H}N_R^0 -y_L\phi \bar{N}_L^{0,c} N_L^0 -y_R\phi \bar{N}_R^{0,c} N_R^0 + {\rm h.c.}\\
    & + \left(D_\mu \phi \right)^\dagger D^\mu \phi  +V(\phi),
\end{aligned}
\end{equation}
where $Z'_{\mu\nu}=\partial_\mu Z'_\nu - \partial_\nu Z'_\mu$ is the field-strength tensor, $H$ is the standard model Higgs doublet, the covariant derivative is given by $D_\mu = \partial_\mu -ig_Y \frac{Y}{2}B_\mu  -ig_W T^iW_\mu^i - ig_{Z'} Y' Z'_\mu$, where $B_\mu$, $W_\mu^i$ and $Z'_\mu$ represent the gauge fields for the ${\rm U(1)}_Y$, ${\rm SU(2)}_L$ and ${\rm U(1)}_{L_\mu-L_\tau}$ groups, respectively. The constants $g_Y$, $g_W$ and $g_{Z'}$ are their associated coupling constants. All fields with a superscript `0' refer to the interacting eigenstates. Furthermore, $L_\alpha$ and $\ell_{\alpha,R}$ represent the SM left-handed and right-handed leptons, respectively, with $\alpha$ denoting their flavor. It is important to note that only $\nu_\mu$ couples to $N_R$, while $\nu_\tau$ does not due to the  ${\rm U(1)}_{L_\mu-L_\tau}$ charge assignment.

\begin{figure}[ht]
    \centering
    \includegraphics[width=0.435\textwidth]{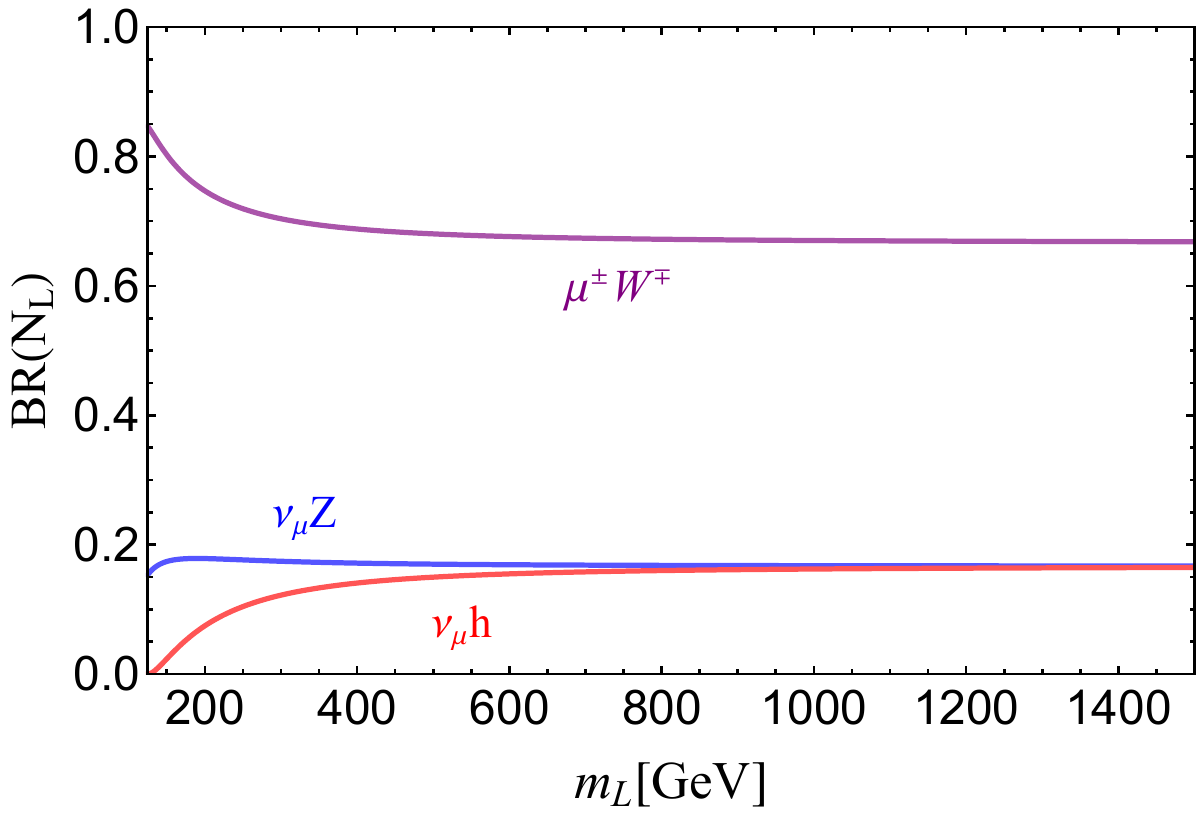}
    \includegraphics[width=0.45\textwidth]{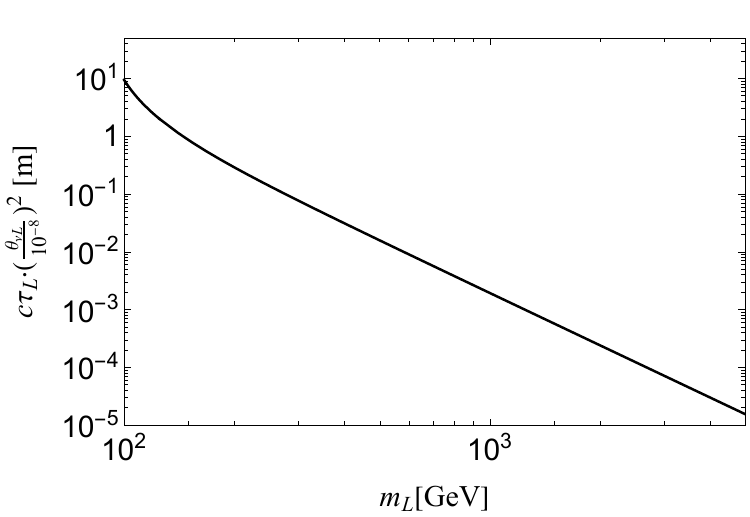}
    \caption{The branching ratios of sterile neutrino $N_L$ decay (left) and its proper decay length (right) as a function of sterile neutrino mass $m_L$. The three main decay branching ratios are shown in different colors with purple corresponding to $N_L\to \mu^\pm W^\mp$, blue corresponding to $N_L\to \nu_\mu Z$, and red corresponding to $N_L\to \nu_\mu h$.}
    \label{fig:br_NL}
\end{figure}

After the electroweak and ${\rm U(1)}_{L_\mu-L_\tau}$ symmetry breaking, the scalar fields are given by $H=\left (\begin{array}{c} 0 \\ v_h+h \end{array}\right)$ and $\phi=v_\varphi+\varphi$, where $v_{h/\varphi}$ denotes the vacuum expectation value (vev) of the Higgs and scalar $\phi$, respectively. Then the effective Lagrangian can be written as
\begin{equation}\label{eq:2}
\begin{aligned}
     \mathcal{L}_{\rm eff}  =&  \mathcal{L}_{\rm SM} -\frac{1}{4}Z'_{\mu\nu}Z'^{\mu\nu} + \frac{1}{2}m_{Z'}^2 Z'_\mu Z'^\mu + \bar{N}^0 i\partial_\mu\gamma^\mu N^0 - m_N \bar{N}^0 N^0 \\
    & +  g_{Z'} Z'_\alpha \left(\bar{\mu}\gamma^\alpha \mu - \bar{\tau}\gamma^\alpha \tau + \bar{\nu}_{\mu,L}^0 \gamma^\alpha \nu_{\mu,L}^0 - \bar{\nu}_{\tau,L} \gamma^\alpha \nu_{\tau,L} + \bar{N}^0\gamma^\alpha N^0\right)\\
    &+\left( - y' \frac{(v_h+h)}{\sqrt{2}} \bar{\nu}_{\mu,L}^0 N_R^0 - y_L \frac{(v_\varphi + \varphi)}{\sqrt{2}} \bar{N}^{0,c}_L N_L^0 - y_R \frac{(v_\varphi + \varphi)}{\sqrt{2}} \bar{N}^{0,c}_R N_R^0 +{\rm h.c.}\right)\\
    &+\frac{1}{2}\partial_\mu\varphi \partial^\mu\varphi-\frac{1}{2}m_\varphi^2 \varphi^2 + 2v_\varphi g_{Z'}^2 Y'^2 \varphi Z'_\mu Z'^\mu + g_{Z'}^2 Y'^2 \varphi^2 Z'_\mu Z'^\mu + V_{\varphi-{\rm self}},
\end{aligned}
\end{equation}
where $m_{Z'} = 2 g_{Z'} v_\varphi$, and $V_{\varphi-{\rm self}}$ denotes the potential terms for the $\varphi$ self-interactions.
The mass terms of the above formula can be organized as
\begin{equation}
\begin{aligned}
    \mathcal{L}_{\rm mass} &\supset - \frac{1}{2}\bar{n}^{0,c} M n^0 + {\rm h.c.}
    = -\frac{1}{2}\bar{n}^c  M_d n + {\rm h.c.},
\end{aligned}
\end{equation}
where $n$ denotes the mass eigenstate and $n^0$ represents the interaction eigenstate. For simplicity, we will later use $\nu_L$ to represent $\nu_{\mu,L}$ throughout the discussion. The complete definitions of $n^0$ and $M$ are:
\begin{align}
    n^0 \equiv 
    \left(\begin{array}{c}
    v_{L}^0 \\
    N_L^0 \\
    N_R^{0,c} 
    \end{array}\right),~
    M \equiv 
    \left(\begin{array}{ccc}
    0 & 0 & m_D \\
    0 & m_L & m_N\\
    m_D & m_N & m_R
    \end{array}\right),
\end{align}
with 
\begin{align}
    m_D = y' v_h/\sqrt{2}, 
    \quad m_L=\sqrt{2} y_L v_\varphi, 
    \quad m_R=\sqrt{2}y_R v_\varphi. 
\end{align}
The orthogonal matrix $U$ can be used to diagonalize the mass matrix, achieving $U^T M U = M_d$, and $n = U^T n^0$.
We assume the mass hierarchy  $m_R \gg m_L , m_D ,  m_N$, the mass matrix $M$ can be diagonalized to the leading order in $m_R^{-1}$ as follows,
\begin{align}
    M_d = 
    \left(\begin{array}{ccc}
    -\frac{m_D^2}{m_R} & 0 & 0 \\
    0 &  m_L - \frac{m_N^2}{m_R}  & 0\\
    0 & 0 & m_R +\frac{m_D^2+m_N^2}{m_R}
    \end{array}\right)+ \mathcal{O}\left(m_R^{-2}\right).
\end{align}
If we define the mixing angles 
\begin{align}
\theta_N \equiv \frac{m_N}{m_R}, ~~~~ \theta_{\nu R} \equiv \frac{m_D}{m_R}, ~~~~
\theta_{\nu L} \equiv \frac{m_N}{m_L} \frac{m_D}{m_R}  = \frac{m_N}{m_L}\theta_{\nu R},
\end{align}
where $\theta_{N} $ representing $N_L$--$N_R$ mixing, $\theta_{\nu R} $ representing active $\nu$--$N_R$ mixing and $\theta_{\nu L} $ representing active $\nu$--$N_L$ mixing. The orthogonal matrix can be simplified by expanding in terms of $m_R^{-1}$,
\begin{equation}
\begin{aligned}
    U \simeq
    \left(\begin{array}{ccc}
    1 & -\theta_{\nu L} & \theta_{\nu R}\\
    \theta_{\nu L} & 1 & \theta_N  \\
    -\theta_{\nu R} & -\theta_N   & 1
    \end{array}\right) - 
    \left(\begin{array}{ccc}
    \frac{1}{2} \left(\theta_{\nu L}^2+\theta_{\nu_R}^2\right) & \theta_N \theta_{\nu R}\left(1 -\frac{\theta_{\nu L}^2}{\theta_N^2}+\frac{\theta_{\nu L}^2}{\theta_{\nu R}^2}\right) & 0 \\
    \theta_{\nu L}^2\left(\frac{\theta_{\nu R}}{\theta_N}-\frac{\theta_N}{\theta_{\nu R}}\right) & \frac{1}{2}\left(\theta_N^2+ \theta_{\nu L}^2\right) & -\theta_N^2 \frac{\theta_{\nu R}}{\theta_{\nu L}} \\
    \theta_N \theta_{\nu L} &\theta_N^2\left(\frac{\theta_{\nu R}}{\theta_{\nu L}} -\frac{\theta_{\nu L}\theta_{\nu R}}{\theta_N^2}\right)  &\frac{1}{2}\left(\theta_N^2 +\theta_{\nu R}^2 \right)
    \end{array}\right) + \mathcal{O}\left(m_R^{-3}\right).
\end{aligned}
\end{equation}
We retain $U$ up to the order of $\mathcal{O}\left(m_R^{-2}\right)$ while deriving the interaction terms of the Lagrangian, as there is a cancellation at the order of $\mathcal{O}\left(m_R^{-1}\right)$ in deriving Eq.~(\ref{eq:Lagrangian_gauge}) for the $Z'-N_L-\nu_L$ vertex.
From the mass matrix $M_d$ above, the masses of the light active neutrino and the two heavy Majorana sterile neutrinos can be derived as
\begin{align}
    m_\nu &\equiv m_1 \simeq \frac{m_D^2}{m_R}=m_R \theta_{\nu R}^2,   \nonumber\\
    m_2 &= m_L - \frac{m_N^2}{m_R} \simeq m_L, \nonumber \\
    m_3 &=  m_R +\frac{m_D^2+m_N^2}{m_R} 
    \simeq m_R,
\end{align}
with the mass hierarchy $m_\nu \ll m_2 \ll m_3$. Due to the approximate equality, we continue using $m_L$ and $m_R$ to represent the physical masses of $N_L$ and $N_R$, respectively.
The minus sign of active neutrino mass has been eliminated by redefining the neutrino field phase. It is important to note that the right-handed sterile neutrino plays a key role in imparting mass to the active neutrino, consistent with the typical seesaw mechanism. In contrast, the left-handed sterile neutrino does not directly contribute to neutrino mass generation. Since neutrino mass is related to $m_D$ and $m_R$, these parameters can be fixed by the neutrino mass $m_\nu$ and mixing angle $\theta_{\nu R}$ :
\begin{align}
m_R=\frac{m_\nu}{\theta_{\nu R}^2}, \quad m_D=\frac{m_\nu}{\theta_{\nu R}} .
\end{align}
Additionally, the parameters $\theta_N$ and $m_N$ can be determined by:
\begin{align}
m_N=\frac{m_L \theta_{\nu L}}{\theta_{\nu R}}, \quad \theta_N=\frac{m_L \theta_{\nu L}}{m_R \theta_{\nu R}} .
\end{align}
Thus, this model has the following relevant parameters:
\begin{align}
\left\{g_{Z^{\prime}}, ~m_{Z^{\prime}},  ~m_{\varphi}, ~m_L, ~\theta_{\nu L}, ~\theta_{\nu R}\right\}
\end{align}
To achieve a small active neutrino mass, $m_\nu = \frac{m_D^2}{m_R}$, without fine-tuning the Higgs Yukawa coupling in the term $\left(y' \bar{L}_\mu^0 \tilde{H} N_R^0 \right)$, we impose the following requirements:
\begin{align}
m_{\nu} \simeq \mathcal{O}~(0.1) ~{\rm eV}, \quad m_D \gtrsim 1 ~ {\rm MeV},
\end{align}
therefore, the $m_R$ should be larger than 10 TeV.
In our study, we focus primarily on the left-handed sterile neutrino $N_L$ at a muon collider with $\sqrt{s} = 3$ or $10$ TeV. A heavy $m_R$ implies a correspondingly large $v_\varphi$, due to the $\mathcal{O}(1)$ Yukawa coupling $y_R$ between $\varphi$ and $N_R$. To prevent the on-shell production of $N_R$ and $\varphi$ at the muon collider, we assume that both $m_\varphi$ and $m_R$ (e.g. $m_R = 100$ TeV) are significantly heavier than the collider's center-of-mass energy. Thus, we could assume $N_R$ and $\varphi$ are effectively decoupled and $\theta_{\nu R}$ will be fixed by the neutrino mass. 
With these in mind, the free parameters of this model will be reduced to fours:
\begin{align}\label{eq:parameters}
    \{g_{Z'},~m_{Z'},~m_L,~ \theta_{\nu L}\}.
\end{align}
Next, we will consider the Lagrangian in the mass eigenstates. We will decompose interaction terms of the Lagrangian into two parts: gauge interactions and scalar interactions.
\begin{align}
\mathcal{L}_{\rm int} \supset\mathcal{L}_{\rm gauge} +\mathcal{L}_{\rm scalar},
\end{align}
where the gauge Lagrangian related to the neutrino can be expanded in terms of $m_R^{-1}$ to the leading order, considering that $m_R$ is very large 
\begin{equation}\label{eq:Lagrangian_gauge}
\begin{aligned}
    \mathcal{L}_{\rm gauge} \supset&~ \left(\frac{g_W}{\sqrt{2}} \bar{v}_L^0 \slashed{W} \mu_L^0  + {\rm h.c.}\right)
    +\frac{g_W}{2\cos{\theta_W}}\bar{v}_L^0 \slashed{Z} \nu_L^0  \\
    &+ g_{Z'} Z'_\mu \left( \bar{\mu}\gamma^\mu \mu + \bar{\nu}_L^0\gamma^\mu \nu_L^0
    +\bar{N}_L^0\gamma^\mu N_L^0  -\bar{N}^{0,c}_R\gamma^\mu N^{0,c}_R \right)\\
    \simeq&~ \frac{g_W}{\sqrt{2}}W_\mu\left( \bar{\nu}_L\gamma^\mu \mu_L
    + \theta_{\nu R} \bar{N}_R^c\gamma^\mu \mu_L
    - \theta_{\nu L} \bar{N}_L\gamma^\mu \mu_L +\mathrm{h.c.}\right)\\
    & + \frac{g_W}{2\cos{\theta_W}}Z_\mu\left[\bar{v}_L \gamma^\mu \nu_L + \theta_{\nu L}^2\bar{N}_L\gamma^\mu N_L + \theta_{\nu R}^2\bar{N}_R^c\gamma^\mu N_R^c\right.\\
    &\left. +\left(\theta_{\nu R} \bar{v}_L \gamma^\mu N_R^c - \theta_{\nu L} \bar{v}_L \gamma^\mu N_L -\theta_{\nu L}\theta_{\nu R} \bar{N}_L\gamma^\mu N_R^c
    + \mathrm{h.c.}\right)\right] \\
    &+ g_{Z'} Z'_\mu \left[ \bar{\mu}\gamma^\mu \mu + \bar{\nu}_L\gamma^\mu \nu_L
    +\bar{N}_L\gamma^\mu N_L  -\bar{N}^c_R\gamma^\mu N^c_R  \right.\\
    &\left. +\left( 2\theta_{\nu R} \bar{\nu}_L\gamma^\mu N^c_{R} + 2\theta_N\bar{N}_L\gamma^\mu N^c_R - 2\theta_N \theta_{\nu R} \bar{\nu}_L\gamma^\mu N_{L} + {\rm h.c.}
     \right)\right].
\end{aligned}
\end{equation}
From the above Lagrangian, the mixing of $N_L$ with the active neutrino, characterized by the angle $\theta_{\nu L}$, can be quite small due to the two-step mixing process: $N_L \to N_R^c \to \nu_L$. As a result, $N_L$ has the potential to be a long-lived particle at the muon collider. Also, the interactions relevant to the SM Higgs and $\varphi$ are reduced to
\begin{equation}
\begin{aligned}
    \mathcal{L}_{\rm scalar} \supset & -\frac{m_R \theta_{\nu R}}{v_h}h \left( \bar{\nu}_L N_R -\theta_{\nu R} \bar{\nu}_L \nu_L^c - \theta_N  \bar{\nu}_L N_L^c -\theta_{\nu L}\bar{N}_L N_R +\theta_N\theta_{\nu L} \bar{N}_L N_L^c +\theta_{\nu R} \bar{N}_R^c N_R +\mathrm{h.c.}\right)\\
    & - \frac{\varphi}{2v_\varphi} \left[m_L \left(  \bar{N}_L^c N_L + 2\theta_{\nu L} \bar{\nu}_L^c N_L + 2\theta_N \bar{N}_L^c N_R^c \right) + m_R \left(\bar{N}_R^c N_R -2\theta_{\nu R} \bar{\nu}_L^c N_R^c -2\theta_N \bar{N}_L^c N_R^c  \right) + \rm{h.c.} \right]\\
    &+\frac{1}{2}\partial_\mu\varphi \partial^\mu\varphi-\frac{1}{2}m_\varphi^2 \varphi^2 + 2v_\varphi g_{Z'}^2 Y'^2 \varphi Z'_\mu Z'^\mu + g_{Z'}^2 Y'^2 \varphi^2 Z'_\mu Z'^\mu + \mathcal{L}_{\varphi-{\rm self}}.
\end{aligned}
\end{equation}
In the above Lagrangian, we omit the full expression of $\mathcal{L}_{\varphi-{\rm self}}$ since we assume that $\varphi$ is too heavy to be produced at the muon collider. Given the center-of-mass energy of the collider, $\sqrt{s} = 3$ or 10 TeV, searching for the on-shell production of $N_R$ is also not feasible. The final relevant interaction terms of the Lagrangian related to $\nu_L$ and $N_L$, up to order $m_R^{-1}$, can be written as:
\begin{equation}\label{eq:17}
\begin{aligned}
    \mathcal{L}_{\rm int} \supset 
    &~ \frac{g_W}{\sqrt{2}}W_\mu\left( \bar{\nu}_L\gamma^\mu \mu_L  - \theta_{\nu L} \bar{N}_L\gamma^\mu \mu_L +\mathrm{h.c.}\right)\\
    &+\frac{g_W}{2\cos{\theta_W}}Z_\mu
    \left(\bar{v}_L \gamma^\mu \nu_L -\theta_{\nu L} \bar{v}_L \gamma^\mu N_L
    + \mathrm{h.c.}\right) \\
    &+g_{Z'}Z'_\mu\left(\bar{\mu}\gamma^\mu \mu + \bar{\nu}_L\gamma^\mu \nu_L +\bar{N}_L\gamma^\mu N_L \right)\\
    & + \left(  \frac{m_\nu }{v_h}h  \bar{\nu}_L \nu_L^c+ \frac{ m_L \theta_{\nu L} }{v_h} h \bar{\nu}_L N_L^c + \rm{h.c.} \right). 
\end{aligned}
\end{equation}
The partial decay widths of $N_L$ based on above Lagrangian in the limit $m_\ell\to 0$ are~\cite{Atre:2009rg}:
\begin{equation}
\begin{aligned}
    &\Gamma(N_L \to \mu^- W^+)=\Gamma(N_L \to \mu^+ W^-) = \frac{\theta_{\nu L}^2 g_W^2}{64 \pi} \frac{(m_L^2-m_W^2)^2(m_L^2+2m_W^2)}{m_L^3 m_W^2}, \\
    &\Gamma(N_L \to \nu_\mu Z) = \frac{\theta_{\nu L}^2 g_W^2}{128 \pi} \frac{(m_L^2-m_Z^2)^2(m_L^2+2m_Z^2)}{m_L^3 m_W^2}, \\
    &\Gamma(N_L \to \nu_\mu h)=  \frac{\theta_{\nu L}^2 g_W^2}{128\pi}
    \frac{(m_L^2-m_h^2)^2}{m_L m_W^2}, \\
    &\Gamma_{N_L} = 2 \Gamma(N_L \to \mu^- W^+) + \Gamma(N_L \to \nu_\mu Z)+\Gamma(N_L \to \nu_\mu h),
    \label{eq:decayWD}
\end{aligned}
\end{equation}
where $\Gamma_{N_L}$ is the total decay width of $N_L$.
In the large $m_L$ limit, the decay width of $N_L$ is proportional to $\theta_{\nu L}^2 m_L^3 / v_h^2$. The branching ratios of $N_L$ are shown in the left panel of Fig.~\ref{fig:br_NL}, where the one labeled by $\mu^\pm W^\mp$ means the sum of the two charge-conjugated channels, $N_L\to \mu^+ W^-$ and $N_L\to \mu^- W^+$.
We focus on $N_L$ with masses greater than 100 GeV, which is the region of interest. When $m_L$ is significantly larger than the Higgs mass, the branching ratios of $N_L$ are approximate: $N_L \to \mu^{ \pm} W^{\mp}$ at 66\%, $N_L \to \nu_\mu Z$ at 17\%, and $N_L \to \nu_\mu h$ at 17\%. Need to mention that the total width of $N_L$ also can be transformed into $\Gamma_{N_L} \propto m_\nu \left(m_L/m_R\right) \left(m_N^2/v_h^2\right)
$. Therefore, we explicitly see the width of $N_L$ is suppressed by small active neutrino mass $m_\nu$ and the small Dirac mass of sterile neutrinos.

When $m_L \gg m_W, m_Z, m_h$, all decay channels of $N_L$ are open, and the sterile neutrino's proper decay length $c \tau_L$ can be directly obtained from the total decay width of $N_L$ :
\begin{align}
    c\tau_L  = \frac{1}{\Gamma_{N_L}} \simeq 2 ~{\rm m} \times \left( \frac{10^{-8}}{\theta_{\nu L}} \right)^2 \left( \frac{100 ~{\rm GeV}}{m_L} \right)^3,
\end{align}
where $g_W=0.65, m_W=80.4~ \mathrm{GeV}$, and $c$ is the speed of light.
The dependence of the lifetime on $m_L$ is illustrated in the right panel of Fig. 1, where the complete formula for $c \tau_L=1 / \Gamma_{N_L}$ from Eq. (19) is used. In the relevant parameter space, for example, with $\theta_{\nu L}=10^{-8}$ and $m_L=100~ \mathrm{GeV}$, the sterile neutrino $N_L$ has a proper decay length of approximately 9.4 meters, because the decay channel $N_L \rightarrow h \nu_\mu$ is forbidden and must be excluded from the calculation. This long-lived signature could potentially be observed at future muon colliders.

\section{Constraints}
\label{sec:constr}

In this section, we discuss the constraints on our model. Regarding collider searches for HNLs at the LHC, two types of searches apply. Long-lived HNL searches, which target $N_L$ than the gauge bosons in the SM, do not constrain our model since we focus on $m_L > 100$ GeV \cite{CMS:2022fut,CMS:2024hik}. Searches for heavier HNLs $N_R$, primarily produced through W-boson s-channel processes, the constraint on the mixing angles between active neutrino and sterile neutrino is $10^{-2}$ \cite{CMS:2015qur,ATLAS:2015gtp,CMS:2018iaf,CMS:2018jxx}, which are outside our region of interest ($\theta_{\nu L} < 10^{-6}$). Therefore, we neglect LHC constraints in this discussion. 
\begin{figure}[htbp]
    \centering
    \includegraphics[width=0.8\textwidth]{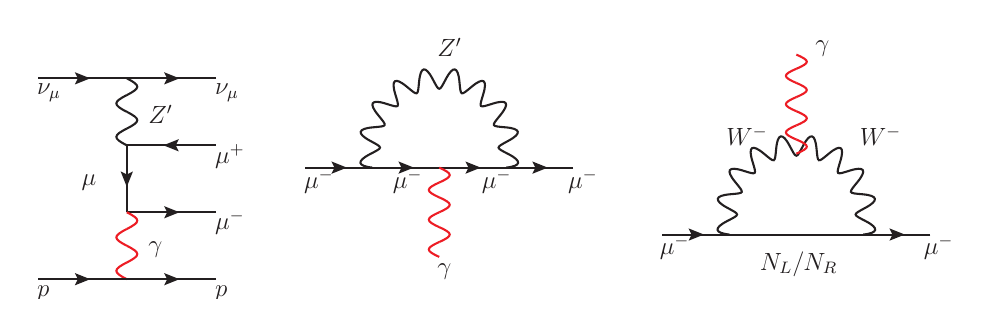}
    \caption{Feynman diagrams of relevant constraint processes for our model. The left panel represents the neutrino trident process, and the middle and right one correspond to the lepton magnetic dipole process.}
    \label{fig:constraints-diagram}
\end{figure}

The ${\rm U(1)}_{L_\mu-L_\tau}$ symmetry imposes natural limits from well-known measurements of the neutrino trident process \cite{Altmannshofer:2014pba}, as shown in the left panel of Fig.~\ref{fig:constraints-diagram}. Introducing a new gauge boson and sterile neutrinos also affects the muon magnetic moment anomaly, depicted in the right panels of Fig.~\ref{fig:constraints-diagram}. We will focus on the constraints from neutrino trident process measurements and the muon magnetic moment anomaly.

The relevant muon-neutrino trident process, illustrated in the left panel of Fig.~\ref{fig:constraints-diagram}, is given by:
\begin{equation}
\nu_\mu ~N \to \nu_\mu ~N ~\mu^+ \mu^-,
\end{equation}
where $N$ represents the nucleus. In our model, the primary contribution to this process comes from the $Z'$ boson. In the heavy mass limit of $m_{Z'} \gg \sqrt{s}$, the total cross section can be expressed as \cite{Altmannshofer:2014cfa}:
\begin{align}
    \sigma({\rm SM}+Z') \approx \frac{1+\left(1+4 \sin^2\theta_W + 2\frac{v^2_h}{(m_{Z'}/g_{Z'})^2} \right)^2}{1+(1+4 \sin^2\theta_W)^2} \sigma({\rm SM}),
\end{align}
where $\sigma({\rm SM})$ is the SM cross section~\cite{Altmannshofer:2014pba}. The inclusive neutrino trident production cross section, $\sigma(\nu_\mu ~N \to \nu_\mu ~N ~\mu^+ \mu^-)$, measured by the CCFR collaboration, agrees well with SM predictions, yielding~\cite{CCFR:1991lpl}:
\begin{align}
    \sigma/\sigma_{\rm SM} = 0.82 \pm 0.28.
\end{align}
Combining the experimental results with theoretical predictions, we obtain the following upper limit (at $95 \%$ C.L.) on the parameters $g_{Z^{\prime}}$ and $m_{Z^{\prime}}$ :
\begin{align}\label{eq:limit_trident}
    \frac{g_{Z'}}{m_{Z'}} < 1.86\times 10^{-3} ~{\rm GeV}^{-1}.
\end{align}
This constraint implies that the vev $v_\varphi$ must exceed $2.69 \times 10^2$ GeV if $m_{Z'} = 2g_{Z'} v_\varphi$. Therefore, the neutrino trident constraints reinforce our earlier assumption of a large $m_R$ and the decoupling of the heavy $\varphi$ and $N_R$ particles.

Another significant constraint comes from the measurement of the muon anomalous magnetic moment $g-2$, which requires $\Delta a_\mu=(249 \pm 48) \times 10^{-11}$ \cite{Muong-2:2023cdq,Borah:2023hqw}. In our framework, the primary contributions to the muon magnetic moment arise from the one-loop diagram involving the ${\rm U(1)}_{L_\mu-L_\tau}$ gauge boson $Z^{\prime}$ (middle panel of Fig.~\ref{fig:constraints-diagram}) and the two sterile neutrinos $N_L$ and $N_R$ (right panel of Fig. \ref{fig:constraints-diagram}). To leading order, the contributions from the $Z^{\prime}$ and sterile neutrino processes can be derived as \cite{Leveille:1977rc}:
\begin{align}
    \Delta a_\mu(Z') &= \frac{ g_{Z'}^2 m^2_\mu}{4\pi^2 m_{Z'}^2} \int_0^1 \frac{x^2(1-x)}{1-x+x^2\frac{m^2_\mu}{m^2_{Z'}}} dx, \nonumber \\
    \Delta a_\mu(N_L) & = \frac{\theta_{\nu L}^2 g_W^2 m^2_\mu}{32\pi^2m_W^2} \int_0^1 \frac{2x^2(1+x) + \frac{m_L^2}{m_W^2}(2x-3x^2+x^3)}{x+\frac{m_L^2}{m_W^2}(1-x)}dx, 
    \nonumber\\
     \Delta a_\mu(N_R) &=\Delta a_\mu(N_L)(m_L\to m_R, \theta_{\nu L}\to \theta_{\nu R}).
\end{align}
In the limits $m_{Z^{\prime}} \gg m_\mu$ and $m_L, m_R \gg m_W$, these formulas can be simplified to:   
\begin{equation}\label{eq:26}
\begin{aligned}
    \Delta a_\mu(Z') &\simeq \frac{g_{Z'}^2 m^2_\mu}{12\pi^2 m_{Z'}^2},\\
    \Delta a_\mu(N_L) & \simeq \frac{G_F}{\sqrt{2}}\frac{\theta_{\nu L}^2 m^2_\mu }{8\pi^2 }f\left(\frac{m_L^2}{m_W^2}\right),  \\
    \Delta a_\mu(N_R) & \simeq \frac{G_F}{\sqrt{2}}\frac{\theta_{\nu R}^2 m^2_\mu }{8\pi^2 }f\left(\frac{m_R^2}{m_W^2}\right),
\end{aligned}
\end{equation}
where $f(r)$ is given by
\begin{align}
    f(r)=\frac{10-43r+78r^2-49r^3+4r^4 +18r^3 \ln(r)}{3(1-r)^4}.
\end{align}
Given that $g_{Z^{\prime}} \gg \theta_{\nu L}, \theta_{\nu R}$ and $m_L, m_R \gg m_\mu, m_W$, the dominant contribution to the muon $g-2$ comes from the $Z^{\prime}$. This places an upper limit on the parameters $g_{Z^{\prime}}$ and $m_{Z^{\prime}}$, constraining the model to ensure compatibility with the observed muon magnetic moment anomaly.
By requiring that the main $g-2$ contribution from the $Z'$ process (the first line of Eq. (\ref{eq:26})) be smaller than $\Delta a_\mu=(249+ 2\times 48)\times10^{-11}$, consistent with the experimental constraint, we can derive the upper bound on the parameters $g_{Z'}$ and $m_{Z'}$ under the assumption that $m_{Z'}\gg m_\mu$:
\begin{equation}
    \frac{g_{Z'}}{m_{Z'}}\lesssim 6.09\times10^{-3}~\rm{GeV^{-1}}.
\end{equation}
Comparing with Eq. (\ref{eq:limit_trident}), we can see that the constraint from the muon $g-2$ is weaker than that from the neutrino trident experiments. 

\section{Production at Future Muon Collider}
\label{sec:prod}

In this section, we will investigate the phenomenology of the sterile neutrino $N_L$ at a future high-energy muon collider. Following the collider settings outlined in Ref. \cite{Accettura:2023ked}, we consider two benchmark scenarios: one with a center-of-mass energy $\sqrt{s}=3$ TeV and an integrated luminosity $\mathcal{L}=1~\mathrm{ab}^{-1}$; and the other with $\sqrt{s}=10$ TeV and $\mathcal{L}=10~\mathrm{ab}^{-1}$.
\begin{figure}[htbp]
    \centering
    \includegraphics[width=0.9\textwidth]{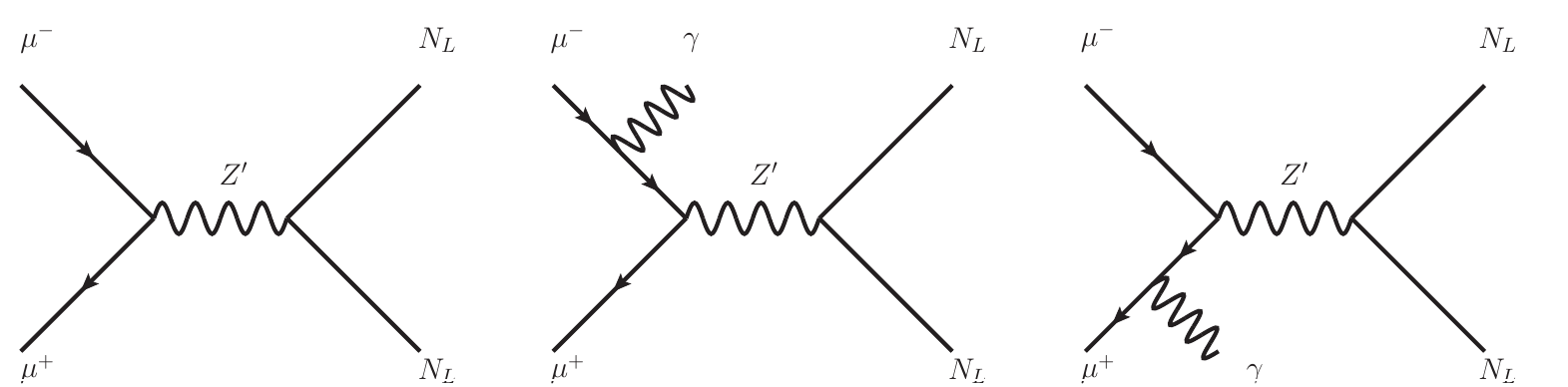}
    \caption{The Feynman diagrams for the production processes of sterile neutrino $N_L$ at the muon collider. The left panel shows the direct pair production mediated by $Z'$, $\mu^+ \mu^- \to N_L N_L$. The middle and the right panels show the production process with an initial radiated photon, $\mu^+ \mu^- \to N_L N_L \gamma$.}
    \label{fig:production-diagram}
\end{figure}

The sterile neutrino $N_L$ can be produced at a muon collider through two main channels based on the interaction of Eq.~(\ref{eq:17}), as depicted in Fig. ~\ref{fig:production-diagram}. The first production mechanism, shown in the left panel, involves direct pair production of $N_L$ mediated by $Z^{\prime}$. The second, shown in the right two panels, involves an initial radiated photon.

\begin{figure}[htbp]
    \centering
    \includegraphics[width=0.49\textwidth]{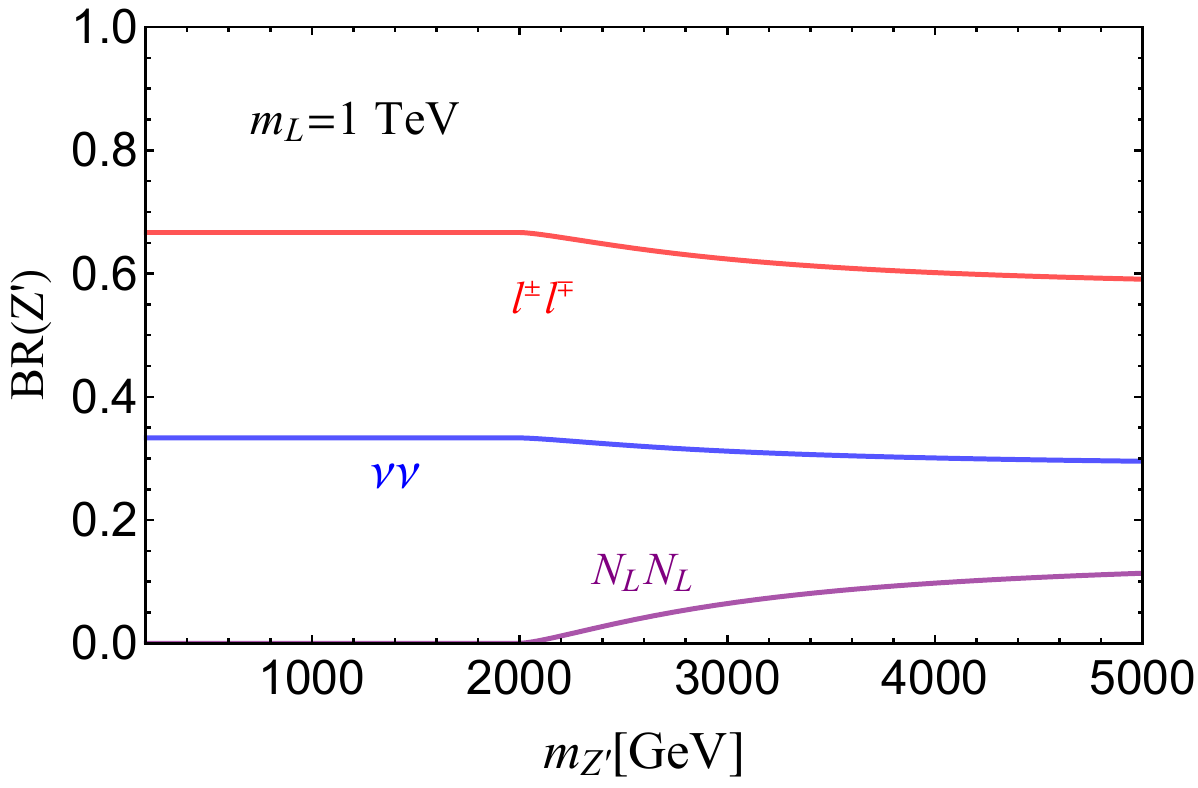}
    \includegraphics[width=0.49\textwidth]{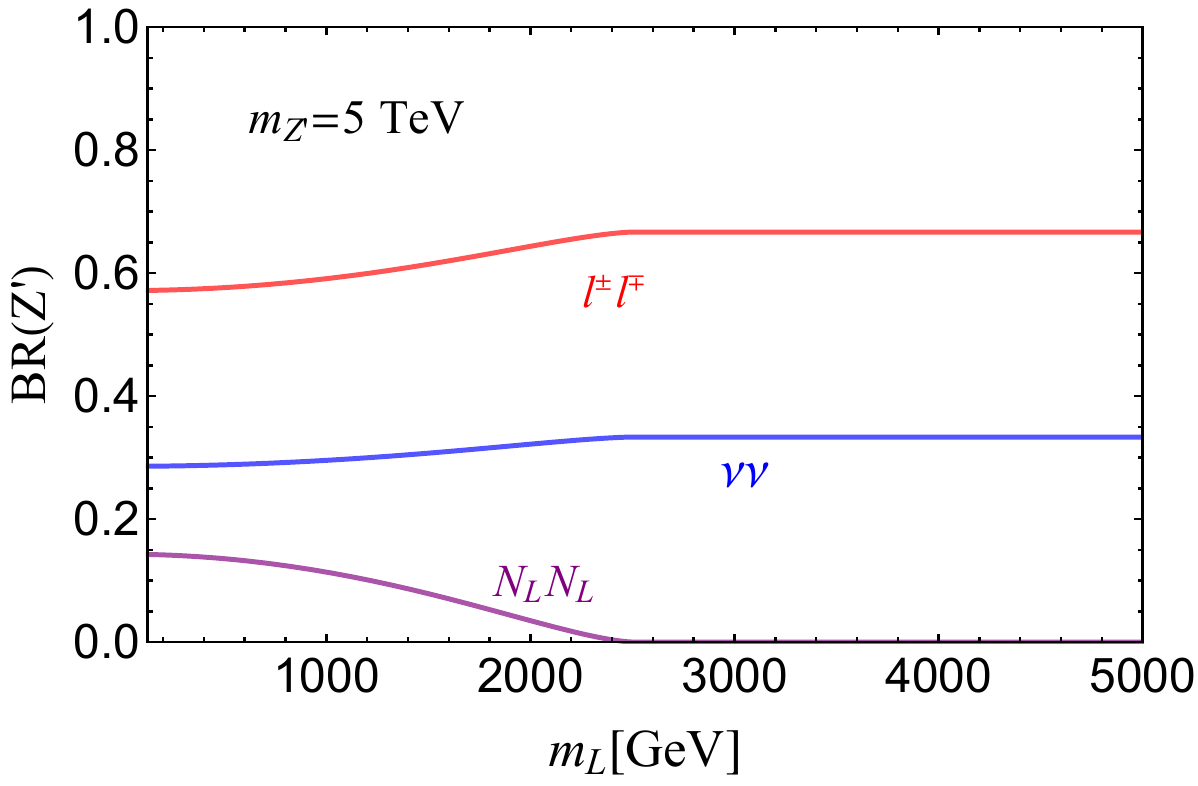}
    \caption{The branching ratios of $Z'$ decays as a function of $m_{Z'}$ (left) and $m_L$ (right). The left (right) panel shows the case with $m_{L}=1$ TeV ($m_{Z'}=5$ TeV), where the red line, blue line, and purple line correspond to the decay channels to charged leptons, neutrinos, and sterile neutrino, respectively.}
    \label{fig:br_Z}
\end{figure}

The pair production of heavy neutral leptons $N_L$ via a $Z'$ boson can occur whenever the $\sqrt{s}>2 m_L$. The cross section for this process is given by:
\begin{align}
    \sigma(\mu^+\mu^- {\longrightarrow}N_L N_L) = \frac{g^4_{Z'}s}{24\pi\left((s-m_{Z'}^2)^2+m_{Z'}^2\Gamma_{Z'}^2\right)}\left(1-\frac{4m_L^2}{s}\right)^{\frac{3}{2}},
\end{align}
where $\Gamma_{Z'}$ is the total decay width of $Z'$, accounting for all possible decay channels, which can be written as:
\begin{align}
\Gamma_{Z'} &= \frac{g^{2}_{Z'}}{12 \pi m_{Z^{\prime}}}\left(\sum_{\ell=\mu, \tau}\left(m_{Z^{\prime}}^2 +2 m_{\ell}^2\right) \sqrt{1-\frac{4 m_{\ell}^2}{m_{Z^{\prime}}^2}} + \sum_{\ell= v_\mu, \nu_\tau} \frac{1}{2}\left(m_{Z^{\prime}}^2 - 4m_{\ell}^2\right) \sqrt{1-\frac{4 m_{\ell}^2}{m_{Z^{\prime}}^2}}\right.\nonumber\\
&\left.+ \frac{1}{2}\left(m_{Z^{\prime}}^2- 4m_L^2\right)\theta(m_{Z'}-2m_L) \sqrt{1-\frac{4 m_L^2}{m_{Z^{\prime}}^2}}\right),
\end{align}
where the first term represents the decay channels for $Z'\to \mu^-\mu^+$ and $Z'\to \tau^-\tau^+$, the second term refers to the channels for $Z'\to \nu_{\mu} \bar{\nu}_{\mu}$ and $Z'\to \nu_{\tau} \bar{\nu}_{\tau}$, and the last term corresponds to the channel for $Z'\to N_L N_L$. Here $\theta(x)$ is the Heaviside step function, which equals 1 for $x>0$ and 0 otherwise.  We show the branching ratios of $Z'$ in Fig. \ref{fig:br_Z} focusing on the case where $m_{L}>100$ GeV. The decay channel $Z'\to N_L N_L$ appears when the mass satisfy $m_{Z'}>2m_L$. When $Z'$ is much heavier than $2 m_L$, the branching ratio among these three channels will satisfy the relation ${\rm BR}(Z'\to \ell \ell):{\rm BR}(Z'\to \nu\nu):{\rm BR}(Z'\to N_LN_L)\simeq 4:2:1$. If $m_{Z'}<2m_L$, the branching ratio will satisfy ${\rm BR}(Z'\to \ell \ell):{\rm BR}(Z'\to \nu\nu):{\rm BR}(Z'\to N_LN_L)\simeq 4:2:0$.

For the production process $\mu^+ \mu^- \to Z'^{(*)} \gamma \to N_L N_L \gamma$ (the middle and right panel of Fig.~\ref{fig:production-diagram}) this also can occur whenever $\sqrt{s}> 2m_L$. The $Z'$ can be either on-shell or off-shell depending on the mass of $Z'$. 
For $\sqrt{s}>m_{Z'}$, $Z'$ can be produced on-shell associated with initial state radiation, which leads to two-body final state $\mu^+ \mu^- \to Z' \gamma$ followed by further decay $Z' \to N_L N_L$ when $m_{Z'}> 2 m_{L}$.  
The cross section in this case can be approximated using the narrow width approximation:
\begin{equation}
    \sigma(\mu^+\mu^- \rightarrow N_L N_L \gamma) \approx \sigma(\mu^+\mu^- \rightarrow Z' \gamma) \cdot {\rm BR}(Z' \rightarrow N_L N_L),
    \label{XS-on-shell}
\end{equation}
where the differential cross section for $Z'\gamma$ production can be calculated as:
\begin{align}
    \frac{d\sigma}{d\cos\theta}(\mu^+\mu^- \rightarrow Z' \gamma) &= \frac{\alpha g^2_{Z'}(1-m_{Z'}^2/s)}{2s\sin^2\theta}\left(1+\cos^2\theta+\frac{4s m_{Z'}^2}{(s-m_{Z'}^2)^2}\right),
\end{align}
where $\theta$ represents the $Z'$ scattering angle with respect to the beam line, and the muon mass is ignored. 
The differential cross section diverges at $\theta \rightarrow 0$ and $\pi$, thus we need the cut on $\theta$ to get a finite cross section. After considering the relationship between the $\theta$ and the pseudorapidity.  In next section for the collider simulation, we choose $p_T^\gamma>20$ GeV and $|\eta_\gamma|<2.5$ to remove the beam-induced backgrounds
typically present at muon colliders. Such pseudorapidity cut implies $\theta \approx 10^{\circ}~(170^{\circ})$, so we choose the integration cut off for the cross section as $\theta \approx$ $10^{\circ}\left(170^{\circ}\right)$. For the parameter choice, $g_{Z'}=0.1$, $m_{Z'}=2$ TeV, and $\sqrt{s}=3$ TeV, the cross section is evaluated as $\sigma(\mu^+\mu^- \rightarrow Z' \gamma) \approx 32.6$ fb. It should be noted that these cross sections were computed numerically using {\tt MadGraph 5}~\cite{Alwall:2014hca}, and are consistent with analytical results of Eq. (\ref{XS-on-shell}).

For $m_{Z'} > \sqrt{s}$, the two-body production $\mu^+ \mu^- \to Z' \gamma$ is kinematically forbidden. In this case, the only viable process is the three-body production $\mu^+ \mu^- \to Z'^{(*)} \gamma \to N_L N_L \gamma$, mediated by an off-shell $Z'$. To account for this contribution, we used {\tt MadGraph 5} to numerically compute the corresponding cross section.
\begin{figure}[htbp]
    \centering
    \includegraphics[width=0.48\textwidth]{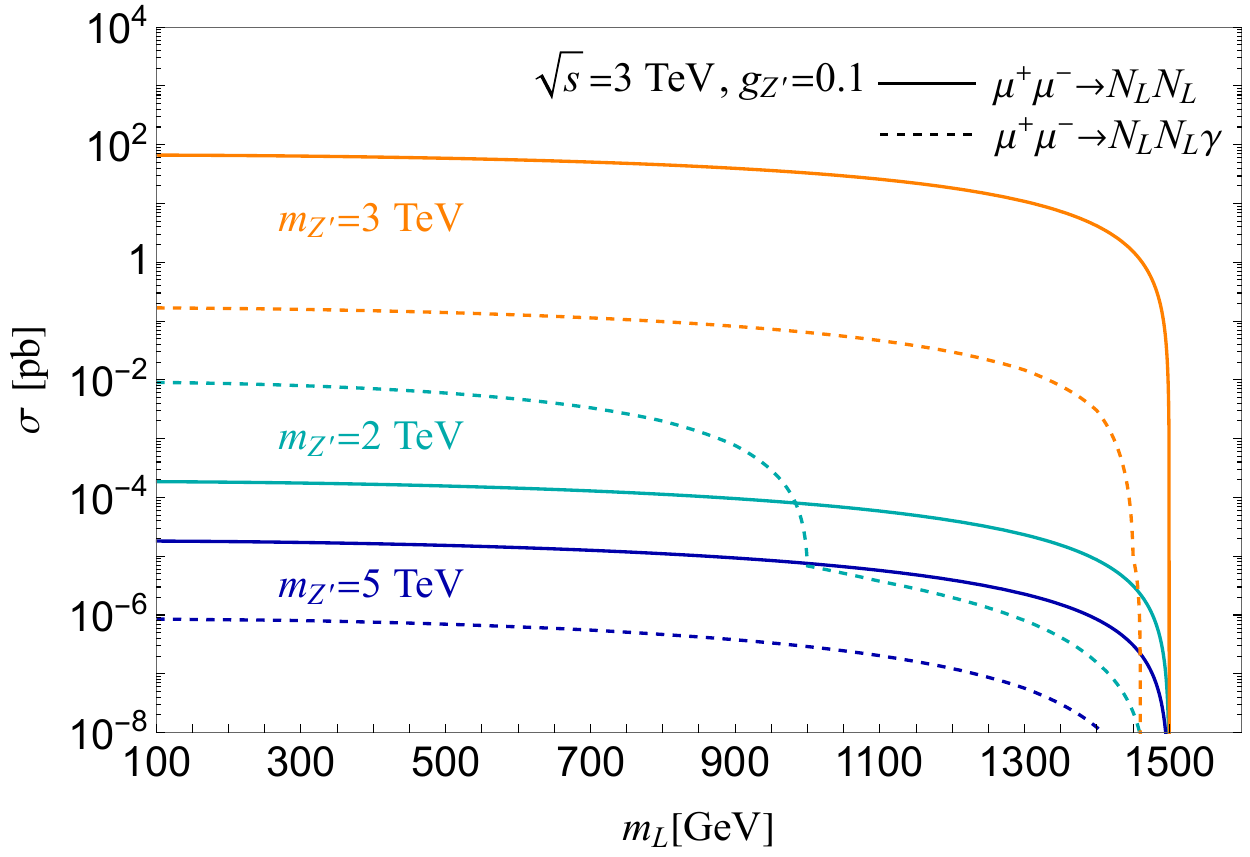}
    \includegraphics[width=0.48\textwidth]{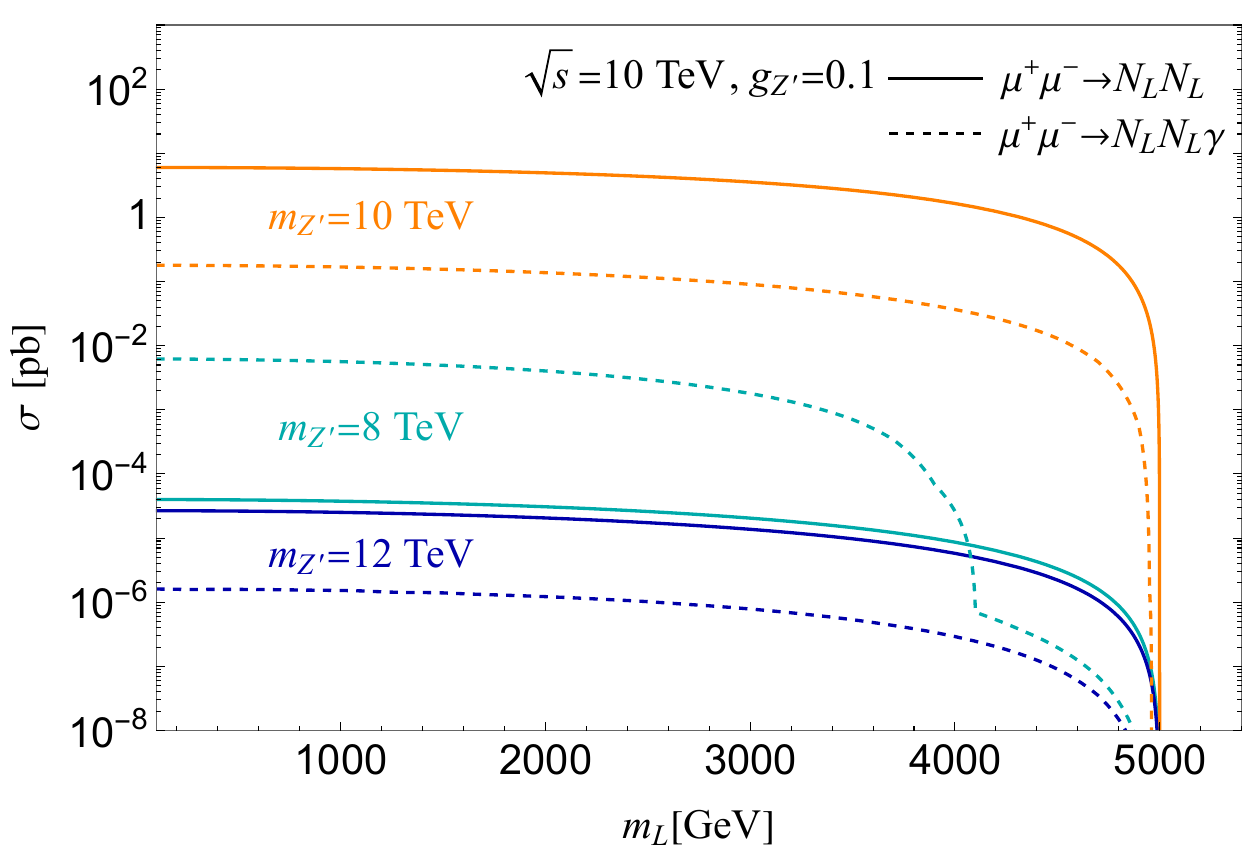}
    \caption{The cross sections for pair production of sterile neutrino $N_L$ are plotted with different center-of-mass energy in the left and the right panels. In each panel, the solid line represents the cross section for the direct production process $\mu^+\mu^-\to N_L N_L$, while the dashed line represents the cross section for the production with initial state radiation $\mu^+\mu^-\to N_L N_L \gamma$. Different masses of $Z'$ are shown in different colors. Besides, the cross section for $\mu^+\mu^-\to N_L N_L$ ($\mu^+\mu^-\to N_L N_L\gamma$) with an on-shell $Z'$ is proportional to $g_{Z'}^2$, while the one for $\mu^+\mu^-\to N_L N_L$ ($\mu^+\mu^-\to N_L N_L\gamma$) with an off-shell $Z'$ is proportional to $g_{Z'}^4$, which shows a different scaling behavior with respect to $g_{Z'}$.}
    \label{fig:production-crossX}
\end{figure}

In order to understand which process is dominant, we show the production cross section in Fig.~\ref{fig:production-crossX} with $g_{Z'}=0.1$, choosing two benchmark center-of-mass energy $\sqrt{s}=3$ TeV and $\sqrt{s}=10$ TeV, respectively. We consider three benchmark values of $m_{Z'}$ for each center-of-mass energy: $m_{Z'}=2$ TeV, $m_{Z'}=3$ TeV, and $m_{Z'}=5$ TeV for $\sqrt{s}=3$ TeV; $m_{Z'}=8$ TeV, $m_{Z'}=10$ TeV, and $m_{Z'}=12$ TeV for $\sqrt{s}=10$ TeV. The solid lines represent the pair production process $\mu^-\mu^+\to N_LN_L$, while the dashed lines indicate the initial state radiation process $\mu^+\mu^-\to N_L N_L \gamma$, which include both $Z'$ on-shell and off-shell. Compare the left and right panels in Fig.~\ref{fig:production-crossX}, when $m_{Z'} \simeq \sqrt{s}$, the total cross section is enhanced, but the $N_LN_L$ process will be enhanced more than the $N_LN_L\gamma$ process, as seen from the comparison of the orange lines with other lines in each plot. If $m_{Z'}< \sqrt{s}$ and $m_{Z'}>2m_L$, the production process with initial state radiation will be dominant. Conversely, if $m_{Z'}> \sqrt{s}$, or $m_{Z'}<2m_L$, the pair production of $N_LN_L$ will be dominant because of the kinematic constraints. In our study, the cross section for $N_L N_L \gamma$ with off-shell $Z'$ is also provided, and we include all processes to perform the inclusive analysis. 

\section{Long-lived Sterile Neutrino Signals at Muon Collider}
\label{sec:pheno}
At the future muon collider, after the pair production of the sterile neutrino $N_L$, including both $N_L N_L$ and $N_L N_L \gamma$, $N_L$ can further decay into $\mu W, \nu Z$, and $\nu h$. With decay widths suppressed by the mixing angle $\theta_{\nu L}$, its proper decay length can be in the meter scale, as shown in Fig. 1. What's more, in the real decay case in the collider, the Lorentz boost factor needs to be taken into account, which makes it testable at lepton colliders through the observation of displaced vertices (DVs).

Because the branching ratio of $N_L \rightarrow \mu W$ is significantly higher than that of the other two channels, which is shown in Fig. \ref{fig:br_NL}, we focus on the processes $N_L$ decays to $\mu W$, and we conduct the inclusive search strategy which requires at least one of the sterile neutrinos to decay inside the designated detector volume. The complete signal process is :
\begin{align}\label{eq:process}
    \mu^+ \mu^- \to (\gamma) N_L N_L, ~N_L\to W^{\pm} \mu^{\mp}, ~W^{\pm} \to jj.
\end{align}
The number of signal events at the muon collider can be expressed as:
\begin{align}
    N = \mathcal{L} \cdot \sigma \cdot \langle \mathbb{P} \cdot \epsilon \rangle,
\end{align}
where $\sigma$ is the cross section for the process Eq.~(\ref{eq:process}), and $\epsilon$ is the kinematical cut efficiency and $\mathbb{P}$ is the probability for one $N_L$ decaying within the designated detector volume. $\langle \mathbb{P} \cdot \epsilon \rangle$ denotes the averaged inclusive efficiency, which can be  calculated event-by-event at the parton level.
\begin{table}[h]
\renewcommand{\arraystretch}{1.2} 
    \centering
    \begin{tabular}{|c|c|c|}
        \hline
        Parameter                                     & $\sqrt{s} = 3$~TeV  &$\sqrt{s} = 10$~TeV \\
        \hline
        Beam momentum [GeV]                           & 1500                & 5000 \\\hline
        Integrated luminosity [$\rm ab^{-1}$]          & 1                   & 10 \\ 
        \hline \hline
        Subsystem                                     & $R$ dimensions [cm]  & $|Z|$ dimensions [cm] \\
        \hline
        Vertex Detector Barrel & 3.0 - 10.4   & 65.0 \\ \hline
        Inner Tracker Barrel   & 12.7 - 55.4                   & 48.2 - 69.2 \\ \hline
        Outer Tracker Barrel   & 81.9 - 148.6 & 124.9 \\
        \hline
    \end{tabular}
    \caption{The muon collider operating scenarios and boundary dimensions of its tracking detector \cite{MuonCollider:2022ded}.}
    \label{tab:muCtracker}
\end{table}

In order to numerically calculate the efficiencies, we first generate the UFO model using {\tt FeynRules} \cite{Alloul:2013bka} based on the sterile neutrino scenario in the ${\rm U(1)}_{L_\mu-L_\tau}$ gauge field, as described in Sec. \ref{sec:model}. Then we simulate signal events at the parton level using {\tt MadGraph 5} \cite{Alwall:2014hca}, by inputting the UFO model. And then the parton level events are passed to the {\tt Pythia8} \cite{Sjostrand:2007gs} and {\tt Delphes} \cite{deFavereau:2013fsa} to generate the showering, hadronization, and detector effects by using the CMS card.

Additionally, the relevant parameters for the muon collider are provided in Tab. \ref{tab:muCtracker}. For a given parton-level event, the probability $\mathbb{P}$ that a long-lived particle $\left(N_L\right)$ decays within the range $\left[r_1 \cdot \hat{\mathbf{r}}, r_2 \cdot \hat{\mathbf{r}}\right]$ along its flight direction $\hat{\mathbf{r}}$ is given by \cite{Cao:2023smj}:
\begin{align}
    \mathbb{P} = {\rm exp}\left(-\frac{r_1}{\gamma \beta c \tau_L}\right) - {\rm exp}\left(-\frac{r_2}{\gamma \beta c \tau_L}\right),
\end{align}
where $\gamma$ is the Lorentz factor of the sterile neutrino $N_L$, $\beta$ is its speed along the $\hat{\mathbf{r}}$ direction, and $\tau_L$ is its proper lifetime.

We require the sterile neutrino's displaced distance $d_L$ along its movement direction $\hat{\mathbf{r}}$ to satisfy $10 \mathrm{~cm}<|d_L \cdot \sin \alpha|<81.9 \mathrm{~cm}$, where $\alpha$ is the angle between the sterile neutrino $N_L$'s momentum direction and the beamline axis. This minimum displacement distance requirement effectively suppresses the SM backgrounds from prompt decays, while the maximum distance requirement ensures a good track reconstruction efficiency, as shown in Tab. \ref{tab:muCtracker}. More precisely, the long-lived sterile neutrinos must decay before reaching the ``Outer Tracker Barrel'', leaving several layers for efficient reconstruction of charged particle tracks. Besides, to further depress the SM background, we also require the invariant mass of reconstructed jets to be around the $W$ boson mass region. With other kinematic requirements, the whole selection criteria can be organized as~\cite{MuonCollider:2022ded}
\begin{equation}\label{eq:DV-cuts}
\begin{aligned}
    {\rm DV}:~~
    &p_T^{\mu}>20 ~{\rm GeV},~|\eta_{\mu}|<2.5,~p_T^j>20 ~{\rm GeV},~|\eta_j|<2.5,\\
    &10~{\rm cm}<|d_L\cdot \sin\alpha|<81.9~{\rm cm},~|d_L\cdot \cos\alpha|<1.25~{\rm m},\\
    &m_{jj}~(m_{j}^{\rm fat})\in [50,~100]~{\rm GeV},
\end{aligned}
\end{equation}
where the particles appearing are required to be well recognized and reconstructed in the Delphes, $p_T$ and $\eta$ are the transverse momentum and pseudorapidity of the muon from $N_L$ decay or the jets from $W$ decay, respectively, to facilitate their identification. 
$m_{jj}~(m_{j}^{\rm fat})$ is the invariant mass of the final hard jets (or fat jet \cite{Heinrich:2014kza}) from $W$ boson decays, where the one closest to the $W$ boson mass is adopted. Since the analysis is performed at the detector level, the isolation between muons and jets is automatically accounted for by the simulation. If the $W$ boson is reconstructed from two hard jets, we require both jets to satisfy the $p_T$, $\eta$ criteria.

The above selection criteria allow the DVs to be well identified, and give negligible SM backgrounds. Exactly speaking, now that the displaced vertex of the signal is reconstructed via a well-reconstructed muon and one (two) fat (energetic) jet(s). With energetic well-reconstructed displaced muon tagging, the main possible SM backgrounds can arise from the interaction of SM particles with the detector materials, QCD hadron and meson events, tau decays, and fake-track coincidental background. Most backgrounds can be reduced to negligible levels by applying an additional invariant mass cut on the displaced jets, where in this paper, the invariant mass $m_{jj}$ ($m_{j}^{\rm fat}$) is required to be around the $W$ boson mass, exactly in the range $[50,~100]$ GeV \cite{He:2024dwh,Cao:2024rzb}. The remaining potential background arises from fake-track coincidental events, where the tracks of muons and jets from SM sources are accidentally mis-reconstructed around the connections of the hits in the tracker system \cite{Liu:2020vur}. In Ref. \cite{Liu:2020vur}, the fake-track background of pure jets is investigated, which shows that this possible background can be eliminated by applying dedicated DV fitting variables cuts, such as $r_{\rm DV}$ (transverse distance between the DV and the origin), $\Delta D_{\rm min}$ (measure of how well the set of candidate tracks fit in a common vertex), $\bar{t}$ (average of the time coordinate of tracks at the fitted DV), $\bar{z}$ (averaged z-coordinate of tracks at the fitted DV) and standard deviations of the time and $z$-coordinates of the constituent tracks at the fitted DV. In this analysis, the existence of muon track can simplify the estimation of this background, because the muons from SM sources are typically prompt, with tracks pointing to the primary interacting point of the muon collider. As a result, with the requirement on the well-reconstructed muon and jet tracks and the invariant mass of dijets or fat jet \cite{Heinrich:2014kza}, the SM backgrounds can be effectively suppressed to a negligible level. A dedicated analysis of these backgrounds is beyond the scope of this work and will be discussed in more detail in future studies.

\begin{figure}[htbp]
    \centering    
    \includegraphics[width=0.475\linewidth]{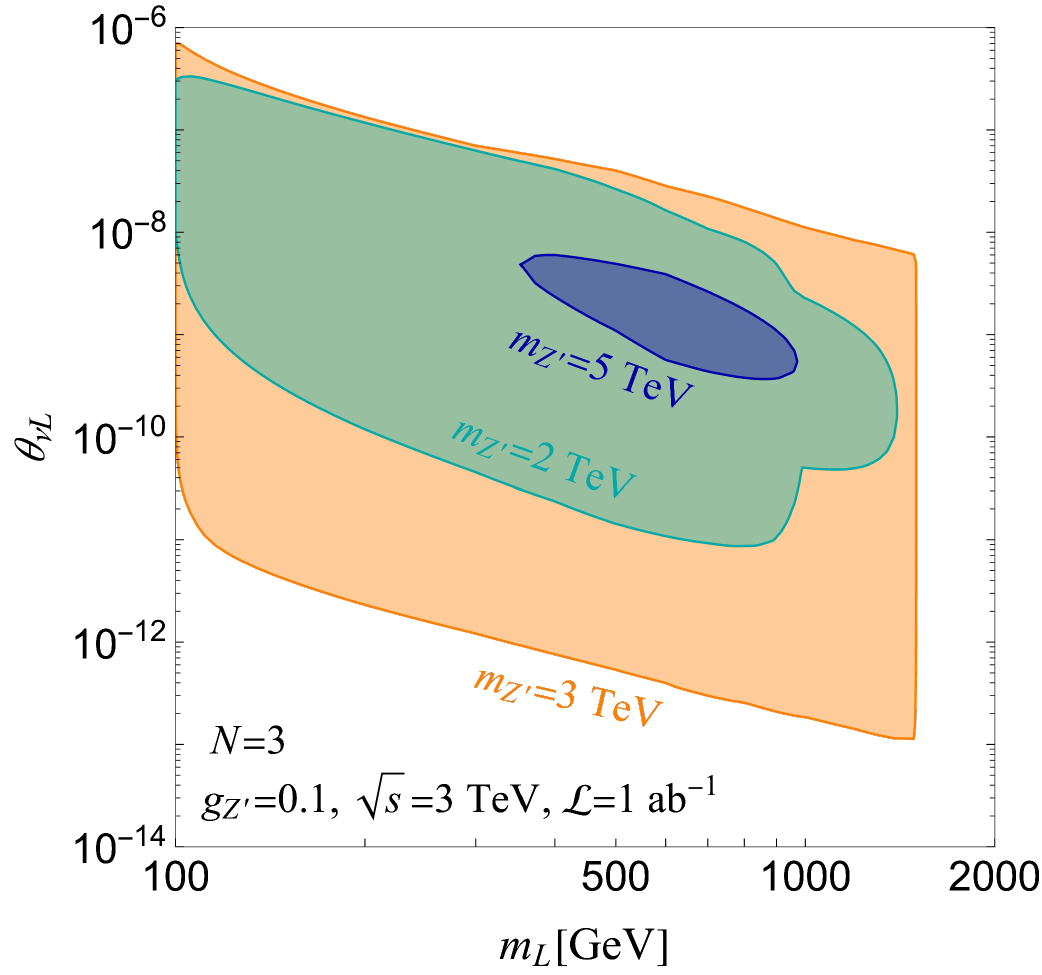}
    \includegraphics[width=0.48\linewidth]{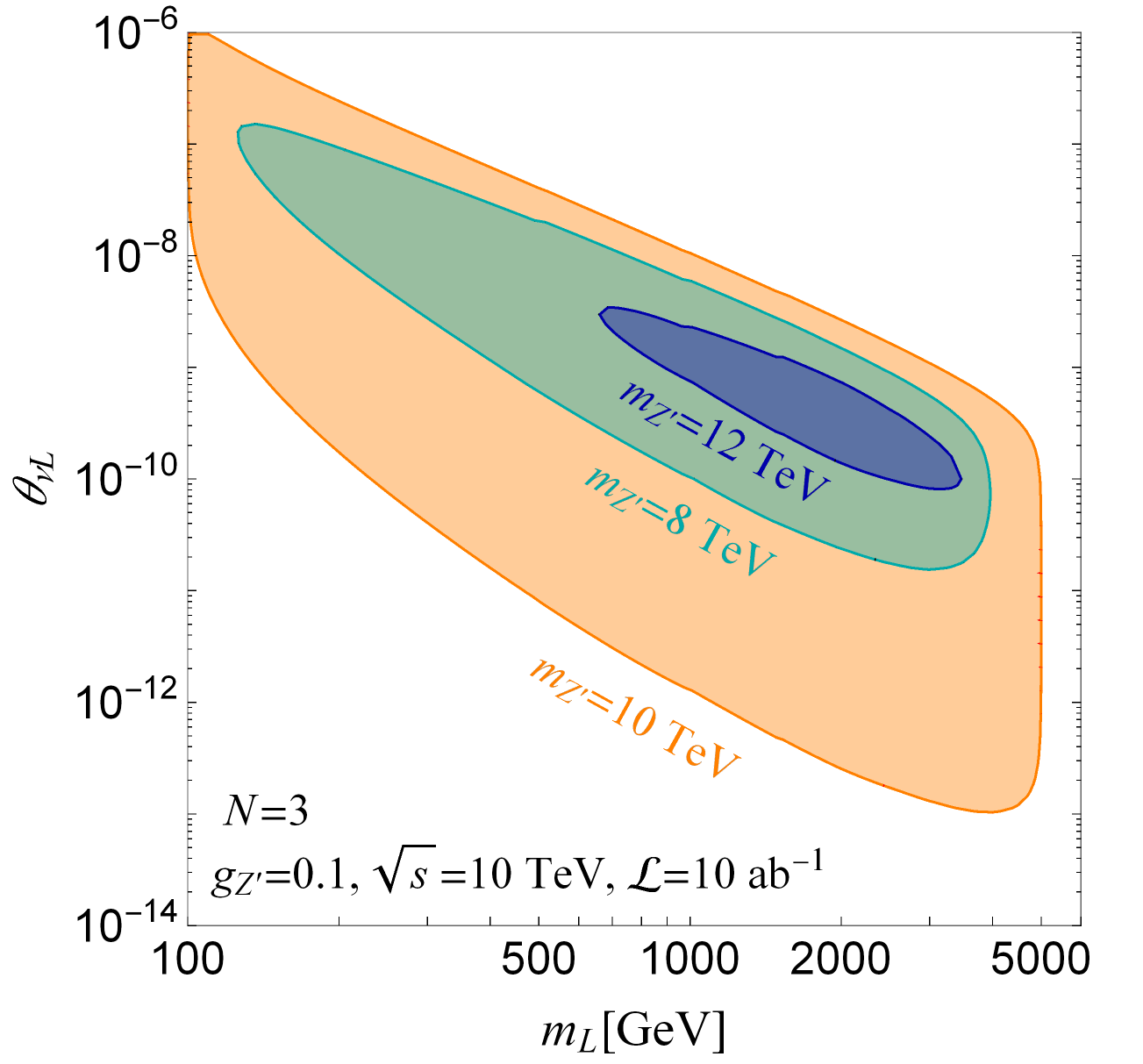}
    \caption{The expected $95\%$ C.L. sensitivities for the inclusive displaced vertex searches for $\mu^+\mu^-\to N_L N_L (\gamma)$ with subsequent decay to $\mu W$ of long-lived sterile neutrino $N_L$ at the muon collider as the function of its mass $m_L$ are shown in the color-shaded regions. The left panel shows the sensitivities for the muon collider with $\mathcal{L}=1~\text{ab}^{-1}$ and $\sqrt{s}=3$ TeV, while the right panel shows the sensitivities for the muon collider with $\mathcal{L}=10~\text{ab}^{-1}$ and $\sqrt{s}=10$ TeV. The orange-shaded region represents the sensitivity with $m_{Z'}=3~(10)$ TeV, the blue-shaded region represents the sensitivity with $m_{Z'}=5~(12)$ TeV, and the cyan-shaded region represent the sensitivity with $m_{Z'}=2~(8)$ TeV combining $\mu^+\mu^-\to N_L N_L$ and $\mu^+\mu^-\to N_L N_L \gamma$ for the left (right) panel.}
    \label{fig:sensi}
\end{figure}

The sensitivities for probing the long-lived $N_L$ at a 3 TeV muon collider with $95\%$ C.L. are shown in the left panel of Fig.~\ref{fig:sensi} with setting the parameter $g_{Z^{\prime}}=0.1$, where the 95\% C.L. corresponds to 3 signal events under the assumption of no SM background. We consider 3 benchmark points chosen for $m_{Z'} = 2, 3, 5$ TeV, which correspond to cyan, orange, and blue-colored regions, respectively. The most sensitive benchmark point is $m_{Z'}= 3$ TeV, because its cross section is enhanced when $\sqrt{s}=m_{Z'}$, which is more contributed from the $N_L$ pair production. For this $m_{Z'}$ choice, the sensitivity for $\theta_{\nu L}$ can be as low as $10^{-13}$. However, the sensitivities decrease rapidly when $m_{Z'}$ moves away from the center of mass energy $\sqrt{s}$, as shown in the left panel of Fig.~\ref{fig:sensi}. When $m_{Z^{\prime}}< \sqrt{s}$ and $m_{Z^{\prime}}>2m_L$, the sensitivity improves due to the higher production cross section of the on-shell $Z^{\prime}$, compared to the off-shell case. Specifically, for $m_{Z^{\prime}}=2$ TeV and $m_L<1000$ GeV, the dominant production process is $Z^{\prime}$ production associated with a photon. The sensitivity is better than that in the case of $m_L>1000$ GeV, where the dominant process is $N_L$ pair production with a sharply reduced cross section. This results in a noticeable kink in the cyan region around $m_L=1000$ GeV, as shown in the left panel of Fig.~\ref{fig:sensi}. Moreover, at $m_{Z^{\prime}}=3$ TeV, the sensitivity exhibits a sharp truncation at $m_L=1500$ GeV not only due to the rapid decline in the production cross section but also because the process becomes kinematically forbidden when $2m_L > \sqrt{s}$. When $Z'$ mass is even heavier, such as $m_{Z'}= 5$ TeV, the production cross section is much smaller due to phase space suppression, resulting in significantly weaker sensitivity compared to the lighter mass cases. 
In all these cases, there is a noticeable decrease in sensitivity for light $m_L$. This reduction of cut efficiency occurs because, when $m_L$ is small, the jets produced from $N_L$ are particularly boosted and collinear, making them difficult to reconstruct independently after processing with \texttt{Pythia8} and \texttt{Delphes}. Moreover, for lighter $m_L$, the issue is more severe because, as $N_L$ becomes highly boosted, reconstructing the muon becomes challenging due to the difficulty in isolating muons from jets. This effect is particularly evident for $m_{Z'} = 5$ TeV, where the sensitivity contour is closed.

The sensitivities for probing the long-lived $N_L$ at a 10 TeV muon collider with $95 \%$ C.L. are plotted in the right panel of Fig.~\ref{fig:sensi}, where we also fixed $g_{Z^{\prime}}=0.1$. The benchmark parameter settings chosen are $m_{Z^{\prime}}=8,10$, and 12 TeV for the cyan, orange, and blue regions in the figure, respectively. The best sensitivity at the muon collider for $\theta_{\nu L}$ can be as good as $10^{-13}$ for $m_{Z^{\prime}}=10$ TeV, where the mass of $Z'$ is equal to $\sqrt{s}$. Away from the resonance mass, the sensitivities decrease significantly due to the reduction in the signal production cross-section. And their shapes are similar to the $\sqrt{s}=3$ TeV case.
On the other hand, the sensitivity on $\theta_{\nu L}$ at 10 TeV is much weaker relative to the 3 TeV muon collider. For example, for $m_L=200$ GeV and $m_{Z'}=3$ (10) TeV, the coupling $\theta_{\nu L}$ in the interval $[3\times10^{-12},10^{-7}]$ ($[2\times10^{-10},3\times10^{-7}]$) can be probed by the collider with $\sqrt{s}=3$ (10) TeV and $\mathcal{L}=1$ (10) $\rm{ab^{-1}}$, where the probing interval of $\sqrt{s}=10$ TeV is somewhat smaller than that of $\sqrt{s}=3$ TeV, as the production cross section varies inversely with the center-of-mass energy $\sqrt{s}$. 
By the way, it is also possible to probe the long-lived sterile neutrino $N_L$ via the time-delayed method due to its slow-moving signatures, if the detector can record the timing information just like the future CMS \cite{CERN-LHCC-2017-027,Contardo:2020886} and ATLAS \cite{Allaire:2018bof} at HL-LHC. In this scenario, the ISR photon with a special transverse momentum cut can also stamp the time of $N_L$ generating. But, conservatively, only the traditional displaced vertex method is applied, and we leave the time-delayed approach in our future work.

\section{Conclusions}
\label{sec:concl}

In this work, we introduce an UV complete model with a Dirac sterile neutrino charged under the ${\rm U(1)}_{L_\mu - L_\tau}$ gauge symmetry. After symmetry breaking, this Dirac sterile neutrino splits into two Majorana sterile neutrinos: one heavy sterile neutrino, $N_R$, which generates the active neutrino mass via a mechanism akin to the type-I seesaw mechanism, and the other naturally long-lived sterile neutrino, $N_L$. The active neutrino mass, induced by mixing with the sterile neutrino $N^c_R$, is governed by the parameters $\theta_{\nu R}$ and $m_R$. The mixing between the left-handed sterile neutrino $N_L$ and the active neutrino $\nu_L$ occurs through a two-step process: $N_L \to N_R^c \to \nu_L$. As a result, the decay width of $N_L$ is doubly suppressed by the smallness of the active neutrino mass and the small Dirac mass of sterile neutrinos, leading $N_L$ to be long-lived in a collider environment.

We explore the long-lived signatures of $N_L$ at future muon colliders, focusing on the inclusive pair production of $N_L$ via $Z'$ exchange in the s-channel, with and without initial photon radiation. Given the decay branching ratio of $N_L$, we concentrate on its subsequent decay into a muon and a $W$ boson. Using the displaced vertex method to detect long-lived $N_L$ decays, we find that at a 3 TeV muon collider with an integrated luminosity of 1 ${\rm ab}^{-1}$, there is significant sensitivity to sterile neutrino masses in the range $m_L \in [100, 1500]$ GeV, with mixing angles $10^{-13} < \theta_{\nu L} < 6 \times 10^{-7}$. At a 10 TeV muon collider with an integrated luminosity of 10 ${\rm ab}^{-1}$, the sensitivity extends to sterile neutrino masses in the range $m_L \in [100, 5000]$ GeV, with mixing angles approximately in the range $10^{-13} < \theta_{\nu L} < 10^{-6}$. These long-lived signatures probe new regions of parameter space for the sterile neutrino and complement other constraints from neutrino trident production experiments and muon $g-2$ measurements.

\section{Acknowledgments}
We thank APCTP, Pohang, Korea, for their hospitality during the focus program [APCTP-2025-F01], from which this work greatly benefited. The work of J. L. is supported by Natural Science Foundation of China under Grants No. 12475103, No. 12235001, and No. 12075005. The work of X. P. W. is supported by the National Science Foundation of China under Grants No. 12375095, and the Fundamental Research Funds for the Central Universities.

\bibliographystyle{JHEP}
\bibliography{ref.bib}
\end{document}